\documentclass[onecolumn,showpacs,preprintnumbers,amsmath,amssymb]{revtex4}
\usepackage{graphicx}% Include figure files
\usepackage{dcolumn}% Align table columns on decimal point
\usepackage{bm}% bold math

\def\code#1{{\tt #1}}

\newcommand{\vecj}{\Vec{j}}
\newcommand{\veck}{\Vec{k}}
\newcommand{\vecr}{\Vec{r}}
\newcommand{\vecx}{\Vec{x}}
\newcommand{\vecy}{\Vec{y}}

\usepackage[
  pdftex,
  pdftitle={An Efficient Algorithm for Classical Density Functional Theory in Three Dimensions: Ionic Solutions},
  pdfauthor={Matthew~G. Knepley, Dmitry~A. Karpeev, Seth Davidovits, Robert~S. Eisenberg, Dirk Gillespie},
  pdfpagemode={UseOutlines},
  bookmarks, bookmarksopen,bookmarksnumbered={True},
  colorlinks, linkcolor={blue},citecolor={blue},urlcolor={blue}
]{hyperref}

\begin{document}

\title{An Efficient Algorithm for Classical Density Functional Theory in Three Dimensions: Ionic Solutions}

\author{Matthew~G. Knepley}
\affiliation{Computation Institute\\University of Chicago}
\email{knepley@ci.uchicago.edu}
\author{Dmitry~A. Karpeev}
\affiliation{Mathematics and Computer Science Division\\Argonne National Laboratory}
\email{karpeev@mcs.anl.gov}
\author{Seth Davidovits}
\affiliation{Department of Applied Physics and Applied Mathematics\\Columbia University}
\author{Robert~S. Eisenberg}
\email{beisenbe@rush.edu}
\author{Dirk Gillespie}
\email{dirk_gillespie@rush.edu}
\affiliation{Department of Molecular Biophysics and Physiology\\Rush University Medical Center}

\date{\today}

\begin{abstract}
Classical density functional theory (DFT) of fluids is a valuable tool to analyze inhomogeneous fluids. However, few
numerical solution algorithms for three-dimensional systems exist. Here we present an efficient numerical scheme for
fluids of charged, hard spheres that uses $\mathcal{O}\left(N\log N\right)$ operations and $\mathcal{O}\left(N\right)$
memory, where $N$ is the number of grid points. This system-size scaling is significant because of the very large $N$
required for three-dimensional systems. The algorithm uses fast Fourier transforms (FFT) to evaluate the convolutions of
the DFT Euler-Lagrange equations and Picard (iterative substitution) iteration with line search to solve the
equations. The pros and cons of this FFT/Picard technique are compared to those of alternative solution methods that use
real-space integration of the convolutions instead of FFTs and Newton iteration instead of Picard. For the hard-sphere
DFT we use Fundamental Measure Theory. For the electrostatic DFT we present two algorithms. One is for the
\textquotedblleft bulk-fluid\textquotedblright\ functional of Rosenfeld [Y. Rosenfeld. \textit{J. Chem. Phys.} 98, 8126
(1993)] that uses $\mathcal{O}\left(N\log N\right)$ operations. The other is for the \textquotedblleft reference fluid
density\textquotedblright\ (RFD) functional [D. Gillespie et al., J. Phys.: Condens. Matter 14, 12129 (2002)]. This
functional is significantly more accurate than the bulk-fluid functional, but the RFD\ algorithm requires
$\mathcal{O}\left(N^{2}\right)$ operations.
\end{abstract}

%\pacs{Valid PACS appear here}

\maketitle
%\keywords{Suggested keywords} % Use showkeys class option if keyword display desired
\section{Introduction}

Since its inception 30 years ago (reviewed by Evans \cite{Evans79}), classical density functional theory (DFT) of fluids
has developed into a fast and accurate theoretical tool to understand the fundamental physics of inhomogeneous
fluids. To determine the structure of a fluid, DFT minimizes a free energy functional
$\Omega\left[\left\{\rho_{k}\left(\vecx\right)\right\}\right]$ by solving the Euler-Lagrange equations
$\delta\Omega/\delta\rho_{k}=0$ for the inhomogeneous density profiles $\rho_{k}\left(\vecx\right)$ of all the particle
species $k$. This approach has been used to model electrolytes, colloids, and charged polymers in confining geometries
and at liquid-vapor interfaces (reviewed by Wu~\cite{Wu06}). Our group has applied one dimensional DFT to biological
problems involving ion channel permeation, successfully matching and predicting experimental
data~\cite{GillespieXuWangMeissner05,Gillespie08}.

DFT is different from direct particle simulations where the trajectories of
many particles are followed over long times to compute averaged quantities of
interest (e.g., density profiles). DFT computes these ensemble-averaged
quantities directly. However, developing an accurate DFT is difficult and not
straightforward. In fact, new, more accurate DFTs are still being developed for
such fundamental systems as hard-sphere fluids
\cite{RothEvansLangKahl02,YuWu02,HansenGoosRoth06}, electrolytes \cite{GillespieNonnerEisenberg02,GillespieNonnerEisenberg03}, and polymers
\cite{YuWu02B}.

When a functional does exist, DFT calculations are, in principle, much faster
than particle simulations because DFT requires solving only a small set of
Euler-Lagrange equations. This is especially true for systems with planar,
spherical, or cylindrical symmetry because in many cases the Euler-Lagrange
equations can be integrated analytically over the extra dimensions. The
resulting equations have only one space variable, while particle simulations
are always performed in three dimensions.

In systems with little or no symmetry, however, the situation is different.
Many of the DFTs for important systems like hard spheres
\cite{Rosenfeld89,Rosenfeld93,RothEvansLangKahl02,YuWu02,HansenGoosRoth06}, Lennard-Jones
dispersion forces \cite{Evans92}, and electrostatic interactions
\cite{Rosenfeld93,GillespieNonnerEisenberg02,GillespieNonnerEisenberg03} require computing a significant number of
convolutions. This increased computational complexity quickly increases
computational time. Moreover, commonly-used numerical techniques scale poorly
with system size, requiring $\mathcal{O}\left(N^2\right)  $ operations (where $N$ is
the number of grid points). For a complex system (e.g., in biology) that
requires $N\gtrsim10^{6}$ for sufficient spatial resolution, this can, in our
experience, mean the difference between 1 week of computer time for an
$\mathcal{O}\left(N^2\right)$ algorithm versus 1 hour for an $\mathcal{O}\left(N\log N\right)$ algorithm. For this
reason, the vast majority of DFT calculations are performed in one dimension, although there are software packages for
three-dimensional system. For example, Tramonto Software for Nanostructured
Fluids in Materials and Biology has been freely available since 2007
\cite{tramontoweb}.

For three-dimensional DFT equations, several different methods are
available to iteratively solve the equations and to evaluate the convolution
integrals. Each choice offers different trade-offs in programming difficulty,
computation time, memory usage, and system size scalability. For example,
Newton iteration requires very few iteration steps compared to Picard
(iterative substitution) iteration, but each Newton step generally takes
significantly longer than a Picard step. For the convolution integrals, either
fast Fourier transforms (FFTs) or real-space methods can be used. FFTs require
a regular, evenly-spaced grid and $\mathcal{O}\left(N\log N\right)$ operations. On
the other hand, real-space methods can (in principle) use an unevenly-spaced
grid (giving a smaller $N$ than required by the FFTs), but require $\mathcal{O}\left(N^2\right)$ operations. The
Tramonto software used Newton iteration with real-space convolution evaluation.

In this paper, we describe a FFT-based Picard iteration method. We chose this approach for several reasons. First, our
numerical experiments showed that Picard iteration was generally faster than Newton and that in systems with liquid-like
concentrations Newton did not always converge. Second, we found that real-space methods are impractical for DFT because
of the specific kernels of the convolution integrals used in DFT. These convolutions integrate the densities
$\rho_{k}\left(\vecx\right)$ over the interiors and surfaces of spheres (described in detail in
Section~\ref{sec:theory}). Neither the sphere interior nor surface can be
represented with sufficient accuracy using real-space methods; however, they can be represented exactly using Fourier
transforms. Lastly, our solution method requires $\mathcal{O}\left(N\log N\right)$ operations and
$\mathcal{O}\left(N\right)$ memory for hard-sphere fluids. Therefore, it scales optimally with system size.

Currently, this optimal scalability is for uncharged hard spheres.
Electrostatics is more complicated. There are two kinds of electrostatic DFTs
in general use, both based upon a perturbation technique. In the
\textquotedblleft bulk-fluid\textquotedblright\ (BF) method, the electrostatic
component of the free energy functional is expanded around a BF
\cite{Rosenfeld93}, while the \textquotedblleft reference fluid
density\textquotedblright\ (RFD) method updates the reference fluid with
information from the ionic densities $\rho_{k}\left(\vecx\right)$
\cite{GillespieNonnerEisenberg02,GillespieNonnerEisenberg03}. The BF method is the most commonly used (in
Tramonto, for example) and we show how to implement it with the optimal
$\mathcal{O}\left(N\log N\right)$ operations and $\mathcal{O}\left(N\right)$ memory
scaling. The BF electrostatic technique can, however, be
\textit{qualitatively} incorrect \cite{GillespieValiskoBoda05} (see Fig.~\ref{fig:table1line2cation}--\ref{fig:table1line2phi}). As we
describe in Section~\ref{sec:rfd}, the mathematical structure of the RFD equations is
fundamentally different from the convolution-based DFTs of hard spheres and
the bulk-fluid electrostatics method. In this paper we also describe an
$\mathcal{O}\left(N^2\right)$ operations and $\mathcal{O}\left(N\right)$ memory
implementation of the RFD electrostatics method. Reducing the number of
operations for the RFD electrostatics method is the subject of future work.

\section{Theory}\label{sec:theory}

    The DFT Euler-Lagrange equations determine the densities $\rho_i(\vecx)$ in equilibrium in the grand canonical
ensemble which is defined by the electrochemical potential for each ion species $i$ in the bath, $\mu^{bath}_i$. The
$\mu^{bath}_i$, in turn, are determined by the bath concentrations $\rho_i^{bath}$, detailed in
Appendix~\ref{app:Calc-of-muBath}. In equilibrium, the flux density for each ion species is identically zero, so that
\begin{equation}
  \nabla\mu_i = 0,
  \label{eq:chemicalPotentialGradient}
\end{equation}
constraining the electrochemical potential for each ion species $\mu_i$ to be a constant, $\mu_i^{bath}$.

    Here the total electrochemical potential $\mu_i(x)$ is a functional of the densities $\rho_i(\vecx)$, which is divided
into three parts, an \emph{external} (ext) potential, an \emph{ideal gas} portion, and an \emph{excess} (ex) chemical
potential:
\begin{equation}
  \mu_i(x) = \mu^{ext}_i\left(\vec{x}\right) + \mu^{ideal}_i\left(\vec{x}\right) + \mu^{ex}_i\left(\vec{x}\right).
  \label{eq:chemicalPotential}
\end{equation}
The ideal gas part is given by
\begin{equation}
  \mu^{ideal}_i\left(\vec{x}\right) = kT \ln\rho_i\left(\vec{x}\right),
  \label{eq:idealChemicalPotential}
\end{equation}
where $\rho_i$ represents the number density of species $i$, $k$ is Boltzmann's constant, and $T$ is the Kelvin
temperature. Moreover, $\mu^{ext}_i$ is the concentration independent part of the electrochemical potential arising from
an external field. We use this to define the problem geometry, such as a hard wall. Lastly, $\mu^{ex}_i$ comes from
particle interactions. Thus, in equilibrium we have
\begin{equation}
  \rho_i\left(\vec{x}\right) = \exp\left(\frac{\mu^{bath}_i
    - \mu_i^{ext}\left(\vec{x}\right)
    - \mu^{ex}_i\left(\vec{x}\right)}{kT}\right).
  \label{eq:rhoEquation}
\end{equation}
This paper outlines an algorithm for Eq.~\ref{eq:rhoEquation} for charged, hard spheres.

For a system of charged hard spheres, DFT decomposes the excess chemical potential into two components, the hard sphere
(HS) and electrostatic (ES) interactions,
\begin{eqnarray}
  \mu^{ex}_i &=& \mu^{HS}_i\left(\vec{x}\right) + \mu^{ES}_i\left(\vec{x}\right) \nonumber\\
             &=& \mu^{HS}_i\left(\vec{x}\right) + \mu^{SC}_i\left(\vec{x}\right) + z_i e \phi\left(\vec{x}\right)
  \label{eq:excessChemicalPotential}
\end{eqnarray}
where the electrostatic component is further decomposed into a mean field contribution, arising from interactions between
uncorrelated ions, and a \emph{screening} (SC) term arising from electrostatic correlations. We define $z_i$ to be the
valence of species $i$ and $e$ the elementary charge. The mean electrostatic potential $\phi$ satisfies Poisson's equation
\begin{equation}
  -\epsilon \Delta\phi = e \sum_i z_i \rho_i\left(\vec{x}\right)
  \label{eq:Poisson}
\end{equation}
where the dielectric coefficient $\epsilon$ is a constant throughout the entire system. % Should we say ``arising from the polarization of the solvent''?
The definition of the hard-sphere and the screening components of $\mu_i$ in terms of $\rho_i$ constitute the heart of
the density functional theory approach and are discussed in detail in subsequent sections.

%We could extend the formulation to handle the case of steady-state flux by replacing the equilibrium condition with the
%Nernst-Planck equations. Following~\cite{GillespieNonnerEisenberg02}, but in three dimensions, the steady state particle
%flux at point $\vec{x}$ can be represented by
%\begin{equation}
%  -\vec{J}_i\left(\vec{x}\right) = \frac{1}{kT} D_i\left(\vec{x}\right) \rho_i\left(\vec{x}\right) \nabla\mu_i\left(\vec{x}\right)
%  \label{eq:diffusion}
%\end{equation}
%along with the continuity equation
%\begin{equation}
%  \nabla\cdot\vec{J_i} = 0
%  \label{eq:continuity}
%\end{equation}
%where the parameter $D_i$ is the diffusion coefficient of species $i$.  The Eqns.~(\ref{eq:diffusion})
%and~(\ref{eq:continuity}), along with Eqn.~(\ref{eq:Poisson}), would normally constitute the Poisson-Nernst-Planck
%system, except that we now allow hard sphere ions.

\section{Hard Sphere Interaction}\label{sec:hardSphere}

The essential DFT-specific modeling of particle interactions is contained in the definition of the chemical potentials
$\mu^{HS}_i$ and $\mu^{ES}_i$. In order to model the interaction of hard spheres, which defines $\mu^{HS}_i$, we use the
Fundamental Measure Theory (FMT)~\cite{Rosenfeld89} developed by Rosenfeld. In FMT, a suitable basis is produced which
best captures the dependence of the potential on the densities. These basis functions, $n_\alpha$, are obtained from
averages of the densities
\begin{equation}
  \label{eq:nDef}
  n_\alpha(\vecx) = \sum_i \int \rho_i(\vecx') \omega^\alpha_i(\vecx' - \vecx) d^3x',
\end{equation}
where the integral is taken over all space and $\alpha \in \{0, 1, 2, 3, V1, V2\}$. The weighting functions
$\omega^{\alpha}_i$ are given by
\begin{align}
  \label{eq:omegaDef}
  \omega^0_i(\vecr) &= \frac{\omega^2_i(\vecr)}{4\pi R^{2}_i} &
    \omega^1_i(\vecr) &= \frac{\omega^2_i(\vecr)}{4\pi R_i} \\
  \omega^2_i(\vecr) &= \delta(|\vecr| - R_i) &
    \omega^3_i(\vecr) &= \theta(|\vecr| - R_i) \nonumber\\
  \vec{\omega}^{V1}_i(\vecr) &= \frac{\vec{\omega}^{V2}_i(\vecr)}{4\pi R_i} &
    \vec{\omega}^{V2}_i(\vecr) &= \frac{\vecr}{|\vecr|}\delta(|\vecr| - R_i) \nonumber
\end{align}
where $\vecr$ is the spherical radial vector. Note that the $V1$ and $V2$ functions are vectors, as are the associated
$n_{V1}$ and $n_{V2}$ functions. If constant concentrations are used in equation~(\ref{eq:nDef}), the ``fundamental
geometric measures'' of the hard spheres (surface area, volume) are recovered.

The HS chemical potential is given by~\cite{Rosenfeld89}
\begin{equation}
  \label{eq:muHS}
  \mu^{HS}_i\left(\vec{x}\right) = kT \sum_\alpha \int \frac{\partial\Phi_{HS}}{\partial n_\alpha}\left(n_\alpha(\vec{x}')\right)
    \omega^\alpha_i\left(\vec{x} - \vec{x}'\right) d^3x'.
\end{equation}
A number of different $\Phi_{HS}(n_\alpha)$ functions have been
developed~\cite{RothEvansLangKahl02,YuWu02,HansenGoosRoth06,Rosenfeld89,Rosenfeld93}, which have different consequences,
most notably the equation of state for a hard sphere fluid modeled with the DFT formalism. We have used the
anti-symmetrized version developed by Rosenfeld et al.~\cite{RosenfeldSchmidtLowenTarazona97},
\begin{align}
  \label{eq:Phi}
  \Phi_{HS}\left(n_{\alpha}\right) &= -n_0 \ln\left(1 - n_3\right) + \frac{n_1 n_2 - \vec{n}_{V1}\cdot\vec{n}_{V2}}{1 - n_3}  \nonumber\\
    &+ \frac{n^3_2}{24\pi \left(1 - n_3\right)^2} \left(1 - \frac{\vec{n}_{V2}\cdot\vec{n}_{V2}}{n^2_2}\right)^3.
\end{align}
However, other choices for $\Phi_{HS}(n_\alpha)$ do not change the numerical scheme we describe below.

It is also important to note that the $n_\alpha$ integrals (\ref{eq:nDef}) are, up to the sign of the argument of the
weight function, convolutions. Since the weight functions $\omega^\alpha$ are either even or odd, we can always convert
the integral to a proper convolution. Therefore, they may be evaluated using the Fourier transform and the convolution
theorem:
\begin{eqnarray}
  n_{\alpha}(\vec{x}) &=& \sum_i \int \rho_i(\vec{x}') \omega^\alpha_i(\vec{x}' - \vec{x}) d^3x' \nonumber\\
                      &=& \mathcal{F}^{-1}\left(\mathcal{F}\left(\rho_i\right) \cdot \mathcal{F}\left(\omega^\alpha_i\right)\right)\nonumber\\
                      &=& \mathcal{F}^{-1}\left(\hat{\rho}_i \cdot \hat{\omega^\alpha_i}\right),
  \label{eq:conv}
\end{eqnarray}
where $\mathcal{F}$ is the Fourier transform operator and the hat denotes the Fourier image of the function. The
chemical potential $\mu^{HS}_i$ can be calculated in exactly the same way, with $\rho_i$ replaced by
$\displaystyle{\frac{\partial\Phi_{HS}}{\partial n_{\alpha}}}$.

In order to evaluate equation~(\ref{eq:conv}), we use the Fast Fourier Transform (FFT) for both the transformation of
$\rho_i$ and the inverse transform of the product $\hat{\rho}_i \cdot \hat{\omega^\alpha_i}$. However, the
$\omega^\alpha_i$ are distributions, and are not easily represented on the rectangular grid required by the FFT. Even a
very fine discretization introduces unacceptably large errors and destroys conservation properties of the basis (e.g.,
conservation of total mass). Thus, if constant concentrations are used in Eq.~\ref{eq:nDef}, the geometric measures of
the sphere are not recovered with straightforward real space methods. In three dimensions, unlike one dimension, in our
numerical experiments these errors persist no  matter how fine a grid is used. This is a severe problem for real space
methods, such as those used in Tramonto. This problem might be resolved through a specialized quadrature, however, the
authors know of no solution yet proposed.

Rather than attempt to discretize the weight functions on a grid, we compute the Fourier transform of each weight
function analytically, and then evaluate them on the same mesh in Fourier space as used by the FFT. The calculations of
the analytic Fourier transforms of $\omega^\alpha$ are given in detail in Appendix~\ref{app:Fourier-of-weights}. This
strategy allows us to calculate machine precision convolutions with arbitrary density fields, whereas the naive
discretization of the weight functions produce substantial errors, often in excess of the field value itself. For
example, using the convolution theorem, Eq.~\ref{eq:conv}, we recover the geometric measures for a constant density
field only when using analytic Fourier Transforms of the weight functions.

\section{Electrostatics}

\subsection{Mean Field}

In order to obtain $\phi$, we solve the Poisson equation~(\ref{eq:Poisson}), for which the source is the charge density
$\sum_i z_i \rho_i$. Since we have access to $\hat{\rho}_i$ from the calculation of $n_{\alpha}$, we may
solve~(\ref{eq:Poisson}) in the Fourier domain, in which the Laplacian is diagonal. Then the mean electrostatic
potential $\phi$ can be calculated by dividing by the eigenvalues of the discrete Fourier transform. At grid vertex
$\vec{j}$, we
have
\begin{equation}
  \hat\phi(\vecj) = \frac{e \sum_i z_i \hat\rho_i(\vecj)}
    {2\epsilon \left(\frac{1 - \cos k_x}{h^2_x} + \frac{1 - \cos k_y}{h^2_y} + \frac{1 - \cos k_z}{h^2_z}\right)}
  \label{eq:potentialSolution}
\end{equation}
where $h_x$, $h_y$, $h_z$ are the grid spacings in each direction, and $k_x$, $k_y$, $k_z$ are calculated as described
in Appendix~\ref{app:Fourier-of-weights}. In order to fully specify the potential, we choose $\hat\phi(0) = 0$, which is
equivalent to having the additional constraint
\begin{equation}
  \int_{\mathcal{D}} \phi = 0.
  \label{eq:potentialConstraint}
\end{equation}
for a domain $\mathcal{D}$.

\subsection{Bulk Fluid Method}\label{sec:bf}

The $\mu^{SC}_i$ component of Eq.~\ref{eq:excessChemicalPotential} attempts to account for electrostatic screening
interactions. In the bulk fluid (BF) model, it is calculated as an expansion around the bath
concentration. From~\cite{GillespieNonnerEisenberg02} we have
\begin{equation}
  \mu^{SC}_i = \mu^{ES,bath}_i - \sum_j \int_{|\vecx-\vecx'| \leq R_{ij}}
    \left(c^{(2)}_{ij}\left(\vecx,\vecx'\right) + \psi_{ij}\left(\vecx,\vecx'\right)\right) \Delta\rho_j(\vecx') d^3x'
  \label{eq:muESexpansion}
\end{equation}
where $R_{ij} = R_i + R_j$, $R_i$ is the radius of ions of species $i$, $\Delta\rho_j = \rho_j - \rho^{bath}_j$,
$c^{(2)}_{ij}\left(\vec{x},\vec{x}'\right)$ is the two-particle direct correlation function,
$\psi_{ij}\left(\vec{x},\vec{x}'\right)$ is the interaction potential of two point particles of charges $z_ie$ and
$z_je$ located at $\vec{x}$ and $\vec{x}'$, so that~\cite{BlumRosenfeld91}
\begin{equation}
  \begin{array}{l}
  c^{(2)}_{ij}\left(\vec{x},\vec{x}'\right) + \psi_{ij}\left(\vec{x},\vec{x}'\right) \\
  \quad = \frac{z_i z_j e^2}{8\pi\epsilon}
    \left(\frac{|\vec{x}-\vec{x}'|}{2\lambda_i\lambda_j} - \frac{\lambda_i + \lambda_j}{\lambda_i\lambda_j} +
    \frac{1}{|\vec{x}-\vec{x}'|} \left(\frac{\left(\lambda_i - \lambda_j\right)^2}{2\lambda_i\lambda_j} + 2\right)\right)
  \end{array}
  \label{eq:CplusPsiFunc}
\end{equation}
where $\lambda_k = R_k + s$ with $s = \frac{1}{2\Gamma}$, the screening length of the
bath~\cite{WaismanLebowitz72,Blum75}. The mean spherical approximation (MSA) screening parameter $\Gamma$ is derived
in~\cite{Blum75} (see also Appendix~\ref{app:Calc-of-muBath}).

The integral in the expansion for $\mu^{SC}_i$ is a convolution, which we also evaluate in the Fourier domain. This
requires $\mathcal{F}(\Delta\rho_j)$, calculated using the FFT, and the transform of equation~(\ref{eq:CplusPsiFunc})
which is calculated analytically below. It should be noted that in this model of electrostatics, transformations of the
$c^{(2)}_{ij} + \psi_{ij}$ need only be calculated once, since they are fixed by the problem parameters. Additionally,
\begin{equation}
  \mathcal{F}\left(\Delta\rho_j\right) = \mathcal{F}\left(\rho_j - \rho_{bath}\right)
    = \mathcal{F}\left(\rho_j\right) - \mathcal{F}\left(\rho_{bath}\right)
\end{equation}
where we have already calculated $\mathcal{F}\left(\rho_j\right)$ for the $n_\alpha$ calculation in Eq.~(\ref{eq:conv}),
and $\mathcal{F}\left(\rho_{bath}\right)$ is a constant. Thus, the only necessary Fourier transform each iteration is
the inverse transformation.

The accuracy of the transform of Eq.~\ref{eq:CplusPsiFunc} is key to the convergence of the nonlinear iteration for the
equilibrium condition. In fact, we were unable to obtain convergence when evaluating these transforms numerically using
the FFT, and were forced to develop analytical expressions. In order to calculate each piece of $\hat c^{(2)}_{ij} +
\hat\psi_{ij}$, we must take the Fourier transform of powers of $r$. The generic term has the form
\begin{equation}
  \int_{{\cal B}(R)} r^n e^{i \vec{k}\cdot\vec{v}}
  = \frac{4\pi}{k} \int^R_0 dr\ r^{n+1} \sin(k r)
  = \frac{4\pi}{k} I_n,
\end{equation}
where $k$ is the magnitude of $\vec{k}$. We derive a recursive definition for the integral $I_n$ using integration
by parts:
\begin{eqnarray}
  I_n = \int^R_0 dr\ r^{n+1} \sin(k r)
  &=& \begin{cases}\left[-\frac{r^{n+1}}{k} \cos(k r)\right]^R_0 + \frac{n+1}{k} J_n & n \ge -1\\
             0 & n < -1\end{cases} \\
  J_n = \int^R_0 dr\ r^n \cos(k r)
  &=& \begin{cases}\left[\frac{r^{n+1}}{k} \sin(k r)\right]^R_0 - \frac{n}{k} J_{n-2} & n \ge 0\\
             0 & n < 0\end{cases}.
\end{eqnarray}
For~Eq.\ref{eq:CplusPsiFunc}, we need the terms
\begin{eqnarray}
  I_{-1} &=& \frac{1}{k} \left(1 - \cos(k R)\right) \\
  I_0    &=& -\frac{R}{k} \cos(k R) + \frac{1}{k^2} \sin(k R) \\
  I_1    &=& -\frac{R^2}{k} \cos(k R) + 2 \frac{R}{k^2} \sin(k R) - \frac{2}{k^3} \left(1 - \cos(k R)\right).
\end{eqnarray}
We also need their limits as $k$ tends to 0,
\begin{eqnarray}
  \lim_{k\to0} \frac{4\pi}{k} I_{-1} &=&  2\pi R^2 \\
  \lim_{k\to0} \frac{4\pi}{k} I_0    &=& \frac{4\pi R^3}{3} \\
  \lim_{k\to0} \frac{4\pi}{k} I_1    &=& \pi R^4.
\end{eqnarray}
Then we have
\begin{equation}
  \hat c^{(2)}_{ij} + \hat\psi_{ij} = \frac{z_i z_j e^2}{\epsilon|\vec{k}|}
    \left(\frac{1}{2\lambda_i\lambda_j} I_1 - \frac{\lambda_i + \lambda_j}{\lambda_i\lambda_j} I_0 +
    \left(\frac{\left(\lambda_i - \lambda_j\right)^2}{2\lambda_i\lambda_j} + 2\right) I_{-1} \right).
  \label{eq:CplusPsiHat}
\end{equation}

\subsection{Reference Fluid Density Method}\label{sec:rfd}

    The Reference Fluid Density (RFD) method is an alternative to the BF method to compute $\mu^{SC}_i$. As shown
in~\cite{GillespieValiskoBoda05} and Fig.~\ref{fig:table1line2cation}--\ref{fig:table1line2phi} below, it
is more accurate than the BF method. The RFD electrostatic functional is detailed
in~\cite{GillespieNonnerEisenberg02,GillespieNonnerEisenberg03}, and briefly summarized here. This perturbation method
approximates $\mu^{SC}_i\left[\left\{\rho_k\left(\vecy\right)\right\}\right]$ with a functional Taylor series, truncated
after the quadratic term, expanded around a reference fluid:
\begin{eqnarray}
  \mu^{SC}_i\left[\left\{\rho_k\left(\vecy\right)\right\}\right] &\approx&
    \mu^{SC}_i\left[\left\{\rho_k^{\mathrm{ref}}\left(\vecy\right)\right\}\right]
    - kT\sum_i \int c_i^{\left(1\right)}\left[\left\{\rho_k^{\mathrm{ref}}\left(\vecy\right)\right\};\vecx\right]
      \Delta\rho_i\left(\vecx\right) d^3x \\
  &-& \frac{kT}{2} \sum_{i,j} \iint
    c_{ij}^{\left(2\right)}\left[\left\{\rho_k^{\mathrm{ref}}\left(\vecy\right)\right\};\vecx,\vecx'\right]
      \Delta\rho_i\left(\vecx\right) \Delta\rho_j\left(\vecx'\right) d^3x\ d^3x'\nonumber
  \label{eq:muESexpansion2}
\end{eqnarray}
with
\begin{equation}
   \Delta\rho_i\left(\vecx\right) = \rho_i\left(\vecx\right) - \rho_i^{\mathrm{ref}}\left(\vecx\right)
\end{equation}
where $\rho_i^{\mathrm{ref}}\left(\vecx\right)$ is a given (and possibly inhomogeneous) reference density profile. By
defining RFD densities to be the bulk densities, we recover the BF perturbation method. The RFD approach
makes the reference fluid densities functionals of the particle densities
$\rho_i\left(\vecx\right)$~\cite{GillespieNonnerEisenberg03}:
\begin{equation}
  \rho_k^{\mathrm{ref}}\left(\vecy\right) = \bar{\rho}_k\left[\left\{\rho_i\left(\vecx\right)\right\};\vecy\right],
  \label{rhoref_rhobar_def}
\end{equation}
where $\bar{\rho}_k$ is the RFD functional. In~\cite{GillespieNonnerEisenberg03}, it is shown that the first-order
direct correlation function (DCF) is given by
\begin{eqnarray}
  c_i^{\left(1\right)}\left(\vecx\right) &=& -\frac{1}{kT} \frac{\delta \mu^{SC}_i}{\delta\rho_i\left(\vecx\right)} \\
  &\approx& \bar{c}_{i}^{\left(1\right)}\left(\vecx\right) + \sum_j \int
    \bar{c}_{ij}^{\left(2\right)}\left(\vecx,\vecx'\right)
    \Delta\rho_j\left(\vecx'\right) d^3x'
  \label{mu_es_bar_eq}
\end{eqnarray}
where
\begin{eqnarray}
  \Delta\rho_k\left(\vecx\right) &=& \rho_{k}\left(\vecx\right) - \bar{\rho}_k\left(\vecx\right),\\
  \bar{c}_i^{\left(1\right)}\left(\vecx\right) &=& c_i^{\left(1\right)}\left[\left\{\bar{\rho}_k\left(\vecy\right)\right\};\vecx\right],\\
  \bar{c}_{ij}^{\left(2\right)}\left(\vecx,\vecx'\right) &=& c_{ij}^{\left(2\right)}\left[\left\{\bar{\rho}_k\left(\vecy\right)\right\};\vecx,\vecx'\right].
\end{eqnarray}

    For the RFD functional, the densities $\bar{\rho}_k\left(\vecx\right)$ must be chosen so that both the first- and
second-order DCFs $\bar{c}_i^{\left(1\right)}$ and $\bar{c}_{ij}^{\left(2\right)}$ can be estimated. This is possible
because the densities $\left\{\bar{\rho}_k\left(\vecx\right)\right\}$ are a mathematical construct and do not represent
a physical fluid. The particular choice of the RFD functional we use here is that of~\cite{GillespieNonnerEisenberg02},
which is also discussed in~\cite{GillespieNonnerEisenberg03}:
\begin{equation}
  \bar{\rho}_i\left[\left\{\rho_k\left(\vecx'\right)\right\};\vecx\right] = \frac{3}{4\pi
    R_{SC}^{3}\left(\vecx\right)}
    \int_{\left\vert\vecx' - \vecx\right\vert \leq R_{SC}\left(\vecx\right)} \alpha_i\left(\vecx'\right)
      \rho_i\left(\vecx'\right) d^3x'
  \label{eq:rhoBarDef}
\end{equation}
where the $\left\{\alpha_k\right\}$ are chosen so that the fluid with densities $\left\{\alpha_{k}\left(\vecx\right)
\rho_k\left(\vecx\right)\right\}$ is charge-neutral and has the same ionic strength as the fluid with densities
$\left\{\rho_k\left(\vecx\right)\right\}$ at every point $\vecx$. The radius of the sphere\ $R_{SC}\left(\vecx\right)$
over which we average is the local electrostatic length scale. Specific formulas for $\alpha_k\left(\vecx\right)$ and
$R_{SC}\left(\vecx\right)$ are given in~\cite{GillespieNonnerEisenberg02,GillespieNonnerEisenberg03}. In order to
estimate the electrostatic DCFs $\bar{c}_i^{\left(1\right)}\left(\vecx\right)$ and
$\bar{c}_{ij}^{\left(2\right)}\left(\vecx,\vecx'\right)$ at each point, we use a bulk formulation (specifically the MSA)
at each point $\vecx$ with densities $\bar{\rho}_k\left(\vecx\right)$, detailed in Appendix~\ref{app:Calc-of-muBath}.

    The RFD reference density $\rho^{ref}(\vecx)$ can be rewritten as the following smoothing operation
\begin{equation}
  \rho^{ref}(\vecx) = \int \rho(\vecx') \frac{\theta\left(|\vecx' - \vecx| - R_{SC}(\vecx)\right)}{\frac{4\pi}{3} R^3_{SC}(\vecx)} dx'
  \label{eq:rhoRef}
\end{equation}
where $\theta\left(x\right) = 1 - H\left(x\right)$ and $H$ is the Heaviside function~\cite{Heaviside}
\begin{equation}
  H\left(x\right) = \begin{cases}0,& x < 0 \\1,& x \ge 0\end{cases}.
\end{equation}
Eq.~\ref{eq:rhoRef} resembles a convolution, but unfortunately the screening radius $R_{SC}(\vecx)$ is nonconstant, and
thus the convolution theorem is inapplicable. We compute $R_{SC}$ using (Eq.~42 in \cite{GillespieNonnerEisenberg02})
\begin{equation}
  R_{SC}(\vecx) = \frac{\sum_i \tilde\rho_i(\vecx) R_i}{\sum_i \tilde\rho_i(\vecx)} + \frac{1}{2\Gamma(\vecx)}
\end{equation}
where $\tilde\rho_i(\vecx)$ indicates the density of species $i$ after we have forced the mixture be locally
electroneutral, but have the same ionic strength.

We can express Eq.~\ref{eq:rhoRef} in the compact notation
\begin{equation}
  \rho^{ref}(\vecx) = \int \mathbb{K}^{\vecx}(\vecx') \rho(\vecx') dx'
  \label{eq:rhoRefKDotProduct}
\end{equation}
where the kernel $\mathbb{K}^{\vecx}(\vecx')$ is given by
\begin{equation}
  \label{eq:screening-kernel}
  \mathbb{K}^{\vecx}(\vecx') = \frac{\theta\left(|\vecx' - \vecx| - R_{SC}(\vecx)\right)}{\frac{4\pi}{3} R^3_{SC}(\vecx)}.
\end{equation}
Since the Fourier transform is an $L_2$ isometry, this expression is equivalent to
\begin{equation}
  \rho^{ref}(\vecx) = \int \left[ \hat{\mathbb{K}}^{\vecx}(\veck) \right]^* \hat\rho(\veck) dk
\end{equation}
where we use the hat to indicate the Fourier transform and star to indicate complex conjugation. Furthermore, we can
calculate the Fourier transform of our kernel analytically. We have
\begin{eqnarray}
  \hat{\mathbb{K}}^{\vecx}(\veck) &=& \int \mathbb{K}^{\vecx}(\vecx') e^{-i \veck\cdot\vecx'} dx' \\
                      &=& \int \frac{\theta\left(|\vecx' - \vecx| - R\right)}{\frac{4\pi}{3} R^3} e^{-i \veck\cdot\vecx'} dx' \\
                      &=& \int \frac{\theta\left(|\vecx''| - R\right)}{\frac{4\pi}{3} R^3} e^{-i \veck\cdot(\vecx'' + \vecx)} dx' \\
                      &=& \frac{3}{4\pi R^3} e^{-i \veck\cdot\vecx} \int^{2\pi}_0 d\phi \int^\pi_0 d\theta \sin\theta \int^R_0 dr r^2 e^{-i \veck\cdot\vecx''} \\
\end{eqnarray}
where $R = R_{SC}(\vecx)$. This integral has been evaluated above in Section~\ref{sec:bf}, so that
\begin{equation}
  \label{eq:screening-kernel-hat}
  \hat{\mathbb{K}}^{\vecx}(\veck) = 3 e^{i \veck\cdot\vecx} \left\{ -\frac{1}{k^2 R^2} \cos kR + \frac{1}{k^3 R^3} \sin kR \right\}.
\end{equation}
Thus we can calculate the action of the screening operator by performing the dot product
in~Eq.~\ref{eq:rhoRefKDotProduct} at each vertex of the real space grid. This algorithm has overall complexity
$\mathcal{O}(N^2)$, however it is accurate to machine precision. Alternative schemes to accelerate the operator
application will be discussed in Section~\ref{sec:conclusions}.

%    In order to check the result, we will compare it against the case of a constant screening radius, to which the
%Fourier transform is applicable,
%\begin{eqnarray}
%  \rho^{ref}(\vecx) &=& \frac{3}{4\pi R^3_{SC}} \int \rho(\vecx') \theta\left(|\vecx' - \vecx| - R_{SC}\right) dx' \\
%                  &=& \frac{3}{4\pi R^3_{SC}} \mathcal{F}^{-1}\left(\int \mathcal{F}(\rho) *
%    \mathcal{F}\left(\theta\left(|\vecx' - \vecx| - R_{SC}\right)\right) \right).
%\end{eqnarray}
%The Fourier transform of the weight function is given by
%\begin{eqnarray}
%  \mathcal{F}(\mathbb{K}) &=& \frac{3}{4\pi R^3_{SC}} \frac{4\pi}{k} \left\{ -\frac{R}{k} \cos kR + \frac{1}{k^2} \sin kR \right\} \\
%                         &=& -\frac{3}{k^2 R^2} \cos kR + \frac{3}{k^3 R^3} \sin kR
%\end{eqnarray}
%and the limit as $k$ approaches zero is one.

In order for this formulation to be consistent, we demand that the screening radius used to construct the reference
fluid density $\rho^{ref}$ is identical to that given by the local MSA closure. Thus, we augment our system of equations
with
\begin{equation}
  \Gamma_{SC}\left[\rho\right](\vecx) = \Gamma_{MSA}\left[\rho^{ref}(\rho)\right](\vecx).
  \label{eq:GammaConsistency}
\end{equation}
Here, the lhs of Eq.~\ref{eq:GammaConsistency} indicates the value of $\Gamma$ used to determine the reference fluid
density using Eq.~\ref{eq:rhoRef}, whereas the rhs is calculated using Eq.~\ref{eq:Gamma}, with $\rho^{ref}$ replacing
$\rho^{bath}$ as the local equilibrium value. This equation is added to our global system at each vertex, producing the
same number of additional equations as another ion species.

\section{Discretization and Solution in Equilibrium}

Problem~(\ref{eq:rhoEquation}) is solved on a rectangular prism domain, supporting a different system size in each
Cartesian direction. This geometry is well supported by the PETSc \code{DA} abstraction~\cite{petsc-user-ref}, which also
allows for easy parallelization. The grid is uniform in each direction, which allows one to compute the convolutions
using Fourier transform techniques. PETSc supports the FFTW package~\cite{FFTW05} automatically. Periodic boundary
conditions are naturally enforced by the FFT.

The bath potential, $\mu^{bath}_i$, and external potential, $\mu^{ext}_i$, are calculated just once during the problem
setup. The geometry is defined using external potentials, $\mu^{ext}$. The excess chemical potential is dependent on the
concentration, as is the electrostatic potential, so these are recalculated at each residual evaluation. Moreover, the
evaluation of the ten $\frac{\partial\Phi_{HS}}{\partial n_{\alpha}}$ and the $n_{\alpha}$ at each grid point must be
done at each residual evaluation since they are also dependent upon $\rho_i$.

\subsection{Nonlinear solver}

Equation~(\ref{eq:rhoEquation}) is a fixed point problem for each ion species $i$,
\begin{equation}
  \rho_i(\vecx) = G\left[\left\{\rho_k(\vecx')\right\}\right].
\end{equation}
The problem is solved using a Picard iteration, since in our experiments Newton's method was both less robust, in that
it did not always converge, and less efficient, since it took more time when it did converge. Each new iterate
$\rho^{(1)}$ is generated from an initial guess $\rho^{(0)}$ using
\begin{equation}
  \rho^{(1)} = G(\rho^{(0)}),
  \label{eq:picard}
\end{equation}
where $\rho$ is understood as a vector of densities over ion species. However, with higher bath densities, it is
necessary to use a line search during the Picard update rather than just successive substitution. Thus, our new guess
$\rho^*$ is given by
\begin{equation}
  \rho^* = (1 - \alpha) \rho^{(1)} + \alpha \rho^{(0)},
\end{equation}
where $\alpha$ is the line search parameter. We determine $\alpha$ by sampling the function $G$ at several densities,
fitting the residual values, $||\rho - G(\rho)||$, to a polynomial in $\alpha$, and choosing $\alpha_\mathrm{min}$
corresponding to the minimum residual value. We currently have a quadratic line search, suggested to us by Roland Roth,
which fits the squared $L_2$ norms of the residuals from Eq.~\ref{eq:picard} as this seemed to better match curves in
the search parameter we sampled for testing.

    In addition, because $n_3(\vecx)$ is the local packing fraction, it should never exceed unity. We bound it by 0.9
which allows us to bound the maximum allowable search parameter $\alpha$ since $n_3$ is a linear function,
\begin{eqnarray*}
  ||n_3((1 - \alpha) \rho^{(0)}) + n_3(\alpha \rho^{(1)})||_\infty &<& ||(1 - \alpha) n_3(\rho^{(0)})||_\infty + ||\alpha n_3(\rho^{(1)})||_\infty \\
                                                           &=& (1 - \alpha) ||n_3(\rho^{(0)})||_\infty + \alpha ||n_3(\rho^{(1)})||_\infty \\
                                                           &<& 0.9.
\end{eqnarray*}
Here, $||\vecx||_\infty$ is the $L^\infty$ norm, which picks the maximum value of $\vecx$ in the finite dimensional
case. This bound was also suggest by Roland Roth. Finally, we have
\begin{equation}
  \alpha < \frac{0.9 - ||n_3(\rho^{(0)})||_\infty}{||n_3(\rho^{(1)})||_\infty - ||n_3(\rho^{(0)})||_\infty}.
\end{equation}

We have also experimented with Newton's method, forming the action of the Jacobian operator using finite
differences. The linear systems are solved with GMRES. Both fixed linear system tolerances and those chosen according to
the Eisenstat-Walker scheme were used. However, the Newton method was not competitive with Picard due to linear
convergence through most Newton steps and the large cost of computing the Jacobian action.

\subsection{Numerical Stability}

    With a coarse grid, there is a potential for serious roundoff error when calculating both the average over an ion
surface, $n_2$, and the directional average, $n_{V2}$. From the definition (\ref{eq:nDef}) we have:
\begin{eqnarray}
  n_2(\vec{x}) &=& \sum_i \int \rho_i(\vec{x}') \omega^2_i(\vec{x} - \vec{x}') d^3x' \\
               &=& \sum_i \int \rho_i(\vec{x}') \delta(|\vec{x} - \vec{x}'| - R_i) d^3x' \\
               &=& \sum_i \int_{\mathcal{S}(R_i)} \rho_i(x + \vec{r}) d\Omega.
\end{eqnarray}
Here $\vec{r} = R_i (\sin\theta\cos\phi, \sin\theta\sin\phi, \cos\theta)$ and $\mathcal{S}(R_i)$ is the surface of a sphere
of radius $R_i$. Likewise, 
\begin{eqnarray}
  n_{V2}(\vec{x}) &=& \sum_i \int \rho_i(\vec{x}') \omega^{V2}_i(\vec{x} - \vec{x}') d^3x' \\
                  &=& \sum_i \int \rho_i(\vec{x}') \frac{\vec{r}}{r} \delta(|\vec{x} - \vec{x}'| - R_i) \\
                  &=& \sum_i \int_{\mathcal{S}(R_i)} \rho_i(\vec{x} + \vec{r}) (\sin\theta\cos\phi, \sin\theta\sin\phi,\cos\theta) d\Omega.
\end{eqnarray}
Appendix~\ref{app:average-bound} shows that
\begin{equation}
  \frac{|n_2|}{|n_{V2}|} \le 1.
\end{equation}
However, discretization errors in the computation of the last term of Eqn.~(\ref{eq:Phi}), or its derivative
$\frac{\partial\Phi}{\partial n_\alpha}$, can combine to produce large values of this ratio, stalling the nonlinear
solve and leading to unphysical artifacts. These artifacts produce large density oscillations at sharp corners along
the geometric boundary. These oscillations eventually cause divergence of the nonlinear iteration and prevent accurate
solution of the equations. We alleviate this problem by enforcing the bound explicitly.

\section{Verification}

    At several points in the calculation, we perform consistency checks of the results. Moreover, we compare our results
to known thermodynamic solutions, in the limit of very fine meshes. We first check that we recover the fundamental measures,
which are easily computed analytically, when we compute $n_\alpha(\rho)$ with a constant unit density. We also check
that in the bath, $n_3$ is equal to the combined volume fraction of the ion bath concentrations.

    Moreover, we can verify that in symmetric situations, such as near a hard wall, the solution must be homogeneous
over each plane parallel to the wall. As described earlier, these consistency checks are satisfied when analytic Fourier
transforms of the weight functions $\omega_i$ are used, but not using the FFT or real space methods. We can solve an
effectively one-dimensional problem with a wall at $z = 0$ and periodic in each dimension in order to compare with
thermodynamic results. With purely hard sphere interactions, we have another consistency check, namely a relation
between the pressure $P$ in the bath and the density of each species at the wall~\cite{Martin88,GillespieValiskoBoda05},
\begin{equation}
  \label{eq:sumRule}
  \beta P = \sum_i \rho_i(R_i),
\end{equation}
where we evaluate the density at the point of closest approach to the wall. The FMT DFT of hard spheres is known to
satisfy Eq.~\ref{eq:sumRule}, but this relation holds only approximately for electrostatic functionals described
here~\cite{GillespieValiskoBoda05}.

\subsection{Hard sphere fluids}

    A very sensitive test for calculations of ionic solutions are thermodynamic sum rules, such as
Eq.~\ref{eq:sumRule}. We use this as the figure of merit to assess the accuracy of our hard sphere calculations.

    A notable advantage of the DFT formulation over particle simulations, such as a Monte Carlo for hard spheres, is
that both very low and very high densities can be handled efficiently with no algorithmic changes. Low densities are
difficult for canonical ensembles, such as canonical MC or Molecular Dynamics (MD), because very large systems are
required for accurate statistics. A grand canonical formulation of MC can mitigate the problems for low densities,
however, high densities still result in jamming and high rejection rates, requiring very long run times. This can
sometimes be repaired using very specialized techniques~\cite{GoodmanSokal89,Frenkel04}, however currently these cannot
be applied to general systems of the type we present below.

\begin{figure}
\begin{center}
\includegraphics[width=7.0in]{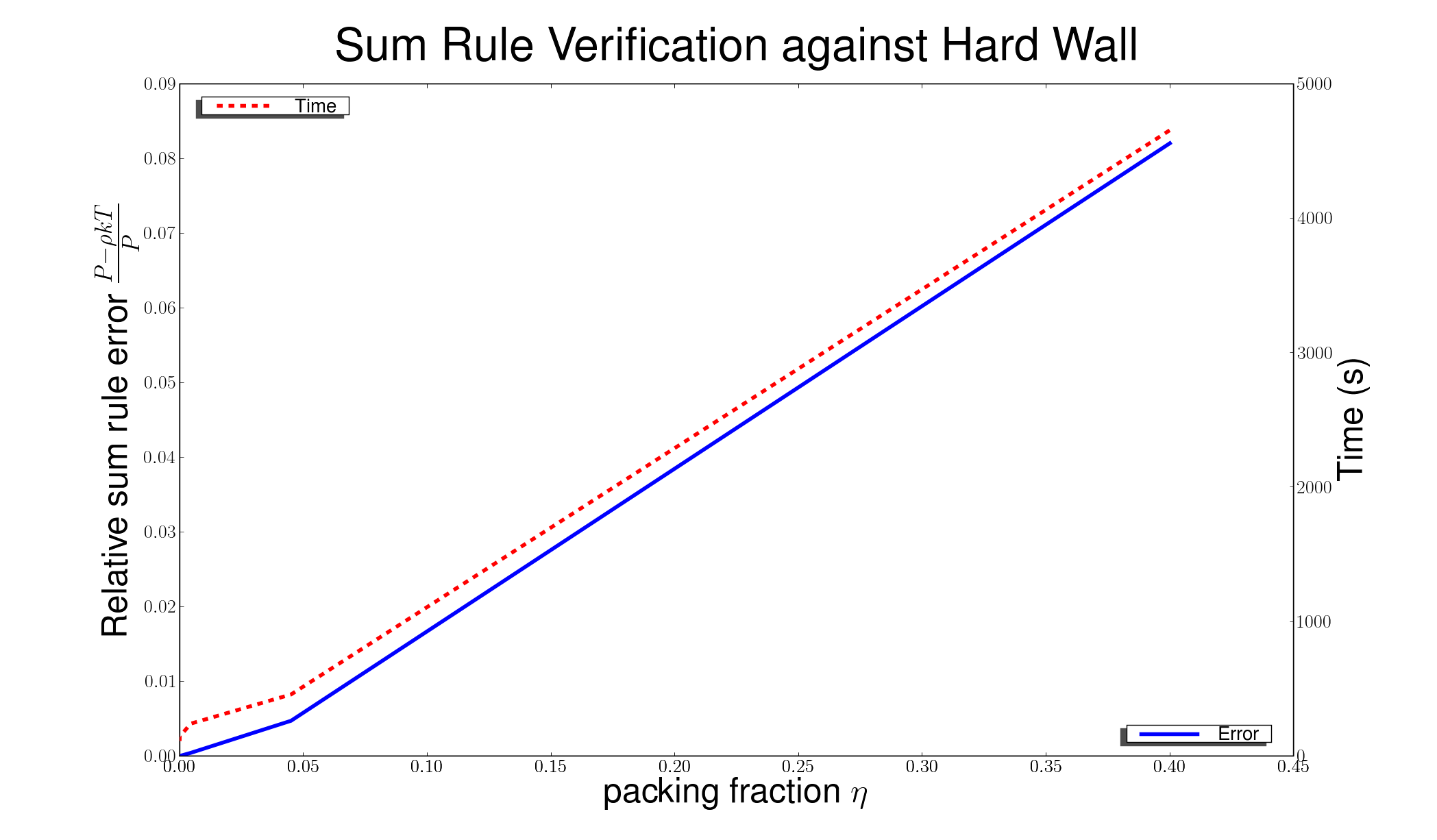}
\end{center}
\caption{Both accuracy and simulation time are shown for a hard sphere liquid of 0.1nm particles. The domain is divided
into cubes which are 0.05nm${}\times{}$0.05nm${}\times{}$.00625nm.}
\label{fig:wall}
\end{figure}

    In order to demonstrate the performance of our algorithm across a range of densities, we simulate a hard sphere
liquid against a hard wall. The particles have radius 0.1nm. In Fig.~\ref{fig:wall}, we show both the simulation time and
accuracy over volume fractions ranging from $10^{-5}$ to $0.4$. Our results are quite accurate, and even at liquid
densities the calculations done on a laptop take less than 1.5 hours. Note that, although this is an effectively one
dimensional problem to facilitate verification, the computation was performed in a full three dimensional geometry.

\subsection{Ionic fluids}

    Calculation of ionic densities near a hard wall also provides a sensitive test for the accuracy of the DFT
method. In~\cite{GillespieValiskoBoda05}, it is demonstrated that the BF version of DFT
(Eq.~\ref{eq:muESexpansion}--\ref{eq:CplusPsiFunc}) provides qualitatively incorrect densities, when compared with the RFD
functional (Eq.~\ref{eq:muESexpansion2}--\ref{eq:rhoBarDef}) and high resolution Monte Carlo simulation. We have successfully
reproduced the one-dimensional DFT and Monte Carlo results with the 3D code, attesting to the accuracy of our
approach. Below, we detail a representative simulation.

\begin{figure}
\begin{center}
\includegraphics[width=7.0in]{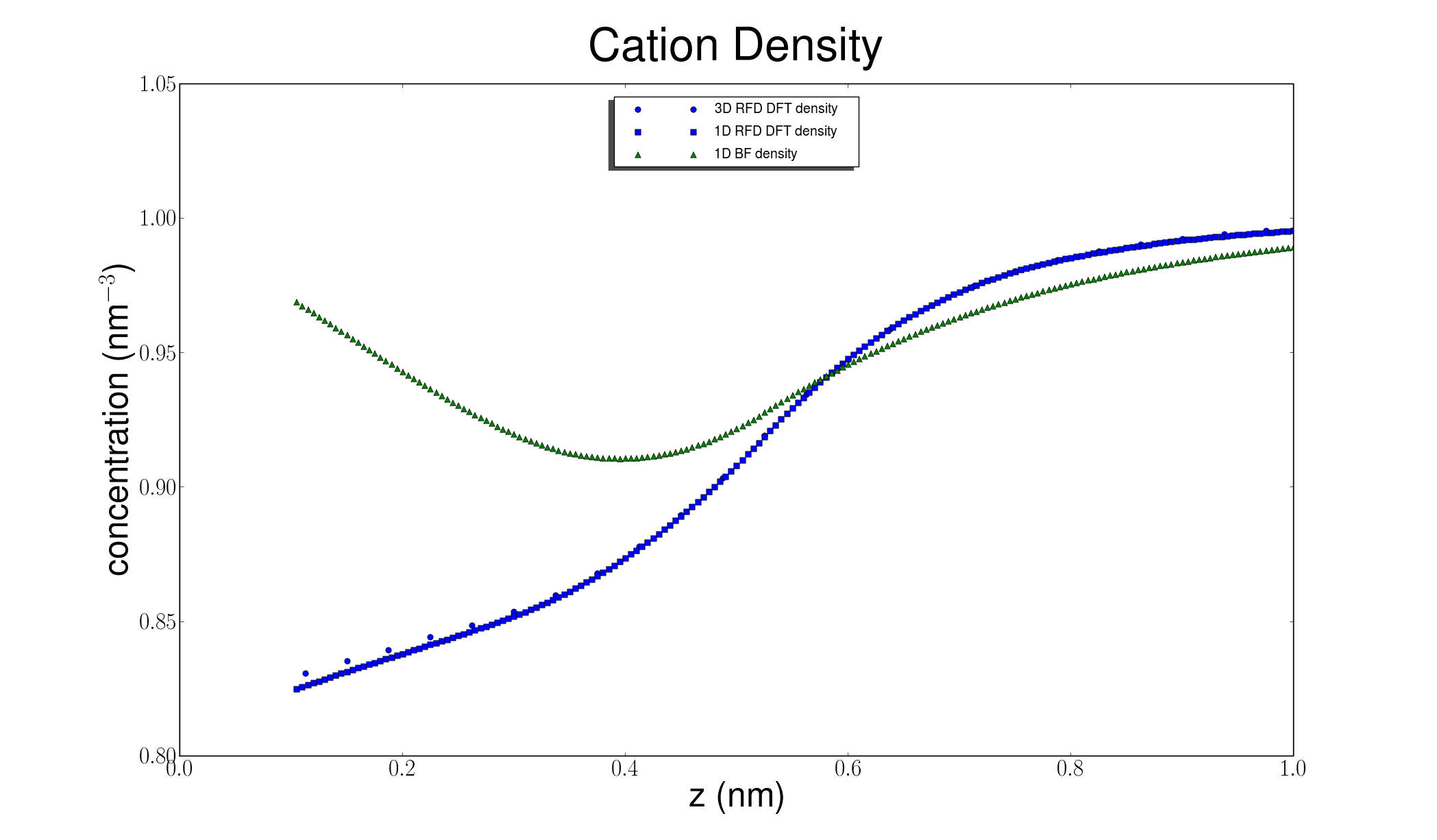}
\end{center}
\caption{Comparing 1D and 3D DFT cation concentrations to MC simulations. The wall is uncharged, the cation
concentration is 1M, and the ions are univalent. The 1D RFD DFT is shown with the blue squares, 3D RFD DFT with blue
circles, and MC with green squares.}
\label{fig:table1line2cation}
\end{figure}

\begin{figure}
\begin{center}
\includegraphics[width=7.0in]{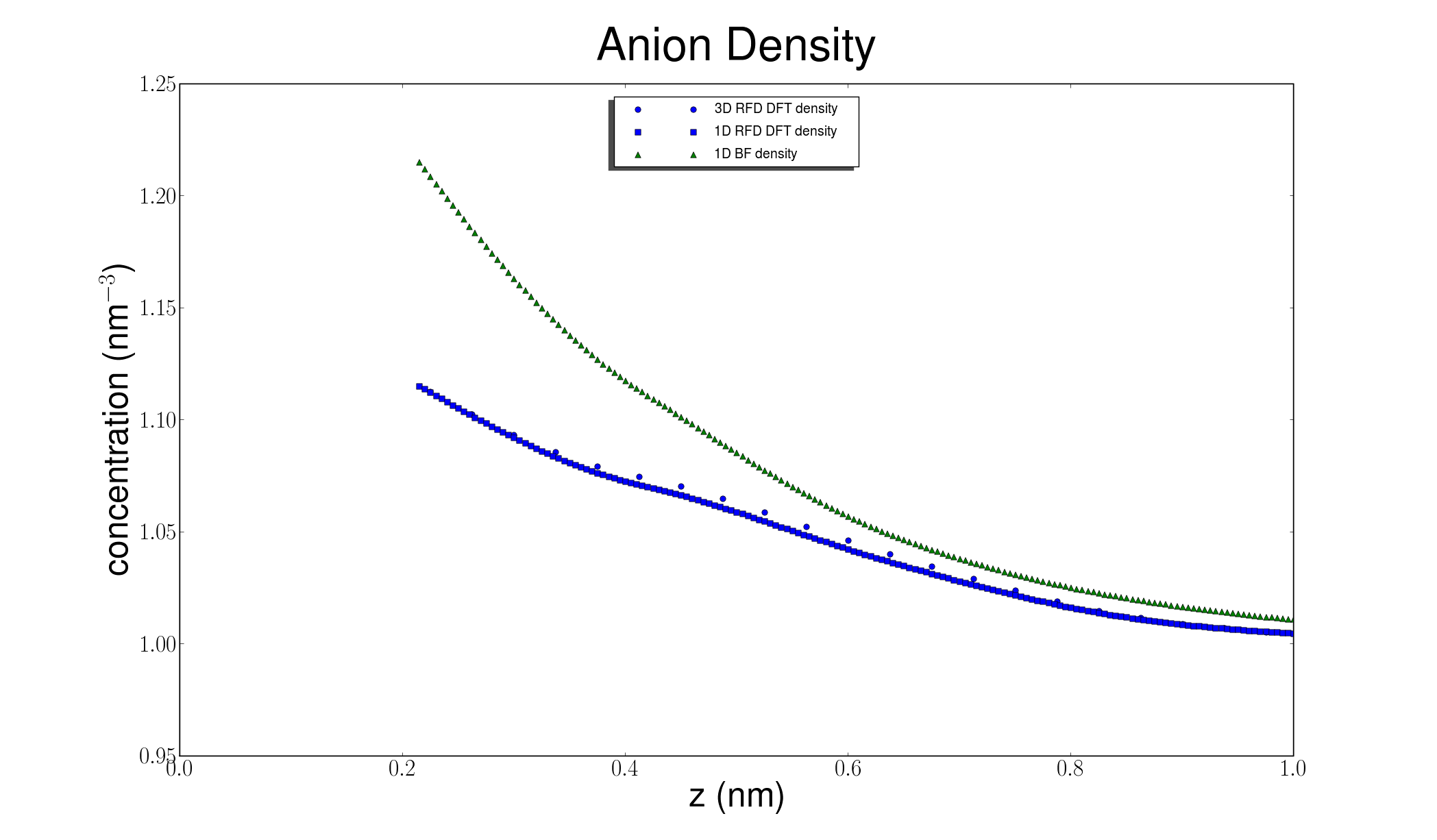}
\end{center}
\caption{Comparing 1D and 3D DFT anion concentrations to MC simulations. The wall is uncharged, the cation
concentration is 1M, and the ions are univalent. The 1D RFD DFT is shown with the blue squares, 3D RFD DFT with blue
circles, and MC with green squares.}
\label{fig:table1line2anion}
\end{figure}

\begin{figure}
\begin{center}
\includegraphics[width=7.0in]{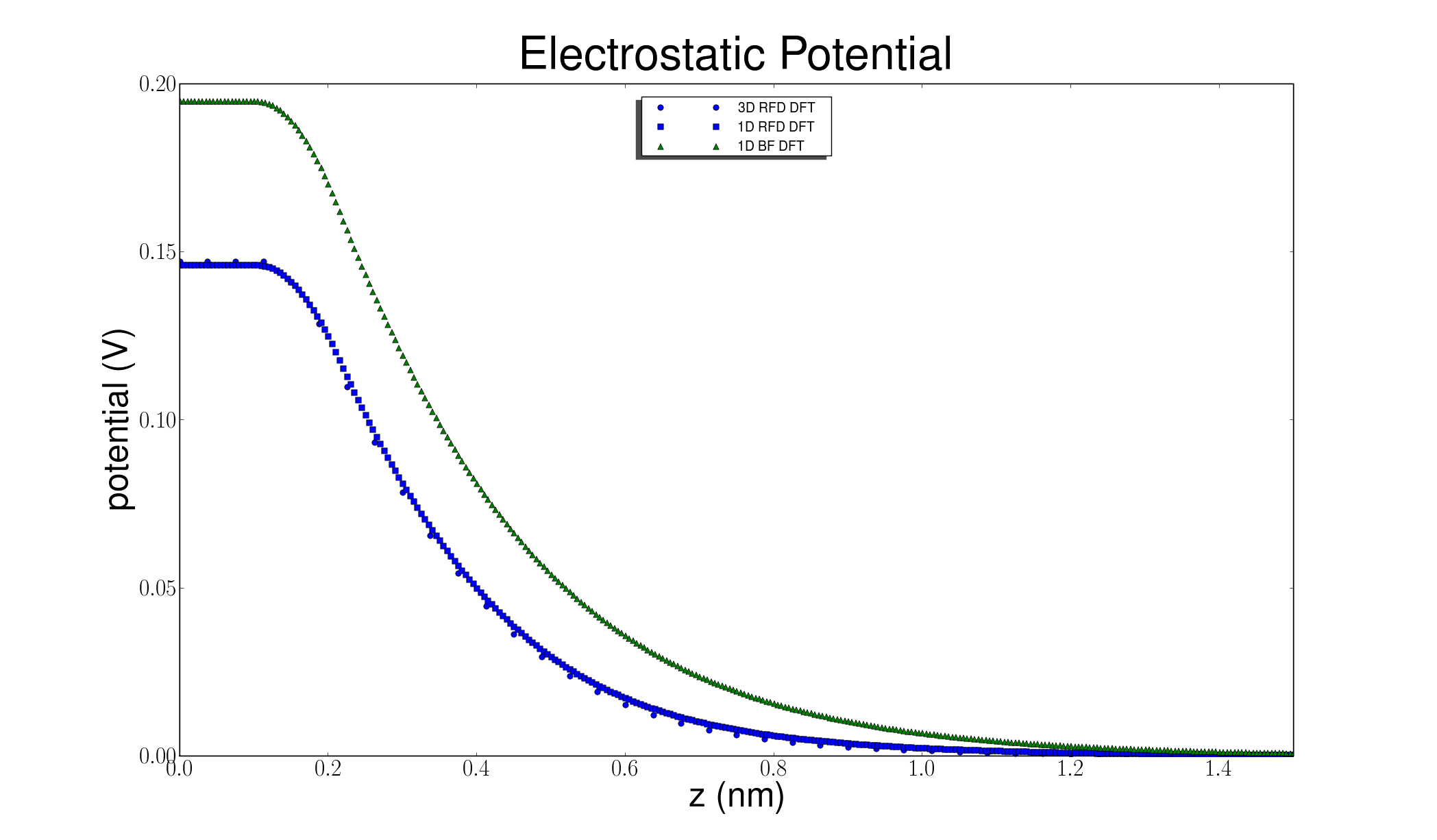}
\end{center}
\caption{Comparing 1D and 3D DFT mean electrostatic potential to MC simulations. The wall is uncharged, the cation
concentration is 1M, and the ions are univalent. The 1D RFD DFT is shown with the blue squares, 3D RFD DFT with blue
circles, and MC with green squares.}
\label{fig:table1line2phi}
\end{figure}

    For our trial calculation, we examine a salt solution of univalent ions. The cation has radius 0.1nm, the anion
0.2125nm. Each species has a 1M bath concentration. The simulation cell, 2nm$\times$2nm$\times$6nm, is periodic in each
direction. A hard, uncharged wall is placed a $z = 0$. We discretize the density on a 21$\times$21$\times$161
grid. The results are insensitive to the resolution in the transverse ($x-y$) directions, but very sensitive in the
normal ($z$) direction. We verify the homogeneity of the solution across $x-y$ planes to machine precision. In
Fig.~\ref{fig:table1line2cation} and Fig.~\ref{fig:table1line2anion}, we show the excellent match between 1D and 3D DFT
results, with MC results shown for comparison. The mean electrostatic potential is shown in
Fig.~\ref{fig:table1line2phi}, also with good agreement.

The BF calculations are currently much more efficient than the RFD calculations, needing only 1.5 minutes compared to
more than a day to run, since BF scales as $\mathcal{O}(N \log N)$, whereas the RFD method scales as
$\mathcal{O}(N^2)$ and additional iterates are needed to obtain a converged reference density. However, the extra
investment of time for RFD computations is necessary because the BF solution is qualitatively incorrect compared to
Monte Carlo simulations.

\section{Conclusions and Future Work}\label{sec:conclusions}

    We have presented a full numerical strategy for solving the three dimensional equilibrium DFT system. The hard
sphere calculation accurately reproduces thermodynamic sum rules, and agrees with prior MC simulation. Moreover,
using the improved RFD electrostatic formulation due to Gillespie et al.~\cite{GillespieNonnerEisenberg02}, we can
accurately reproduce electrostatic behavior near a hard wall for species of differing radii. Thus, the DFT can now
become a powerful tool for full three dimensional chemical simulation, accurately capturing both the energetic and
entropic contributions to the solution.

    There are also several avenues for improvement of the RFD algorithm and extension of the capabilities of the current
code. The dominant cost of this algorithm is the calculation of the reference density used to describe electrostatic
screening. The current algorithm is very accurate, but requires $\mathcal{O}(N^2)$ work. Since the Fourier kernel is
smooth and has rapid decay, it should be possible to construct a multiresolution analysis of it, resulting in a fast
method for application. Moreover, the many FFTs performed at each Picard step could be
replaced by Unequally-Spaced FFTs or wavelet decompositions, which would allow adaptive refinement and increase the size
of problems we can efficiently compute. The FFT and Fast Wavelet Transform (FWT) lend themselves readily to a scalable
parallel implementations. In fact, it should also be possible to offload these transforms onto a multicore co-processor,
such as the Tesla 1060C GPU~\cite{NVIDIATesla}. This will make large scale simulations of charged hard spheres
accessible to working scientists even on a laptop or desktop computer. These algorithmic improvements are the focus of
current research.

%We will also introduce a multilevel formulation and a grid sequencing technique which should accelerate the system
%solution. It is possible that a multigrid formulation will allow the use of Newton's method in finer grids since the
%initial guess may be sufficiently accurate.

\begin{acknowledgments}
This material is based upon work supported by, or in part by, the U. S. Army Research Laboratory and the U. S. Army
Research Office under contract/grant number W911NF-09-1-0488 (DG). The work was also supported by NIH grant GM076013
(RSE).
\end{acknowledgments}

\appendix
\section{Calculation of the Bath Chemical Potential}\label{app:Calc-of-muBath}

    Here we describe the formulas for the electrochemical potential in a homogeneous fluid. When the DFT for hard
spheres uses a Percus-Yevick equation of state~\cite{Lebowitz64}, and the electrostatics is described using
MSA~\cite{WaismanLebowitz72,Blum75}. We follow the treatment in~\cite{NonnerCatacuzzenoEisenberg00}. The bath chemical
potential $\mu^{bath}_i$ has two components, hard sphere and electrostatic,
\begin{equation}
  \mu^{bath}_i = \mu^{HS,bath}_i + \mu^{ES,bath}_i
\end{equation}
which are calculated thermodynamically~\cite{NonnerGillespieHendersonEisenberg01},
\begin{equation}
  \mu^{HS,bath}_i = kT \left( -\ln\Delta + \frac{3\xi_{2}\sigma_i + 3\xi_{1}\sigma_i^{2}}{\Delta}
    + \frac{9\xi_{2}^{2}\sigma_i^{2}}{2\Delta^{2}} + \frac{\pi P^{HS}_{bath}\sigma_i^{3}}{6kT} \right)
  \label{eq:muHSBath}
\end{equation}
based upon the auxiliary variables, where $\sigma_i$ is the ion diameter of species $i$,
\begin{eqnarray}
  \xi_{n} &=& \frac{\pi}{6} \sum_{j} \rho^{bath}_j \sigma_{j}^{n} \qquad n \in \{0,\ldots,3\}\\
  \Delta  &=& 1-\xi_{3}
\end{eqnarray}
and the pressure due to hard sphere interaction in the bath,
\begin{equation}
  P_{bath}^{HS} = \frac{6kT}{\pi} \left( \frac{\xi_{0}}{\Delta} + \frac{3\xi_{1}\xi_{2}}{\Delta^{2}}
    + \frac{3\xi_{2}^{3}}{\Delta^{3}} \right).
  \label{eq:HSPressure}
\end{equation}
The calculation of $\mu_i^{HS,bath}$ as given above is straightforward, but $\mu_i^{ES,bath}$, on the other hand, is
dependent on an implicitly defined parameter $\Gamma$, the MSA inverse screening length,
\begin{equation}
  4 \Gamma^2 = \frac{e^2}{kT \epsilon} \sum_j \rho^{bath}_j \left( \frac{z_j - \eta \sigma^2_j}{1 + \Gamma \sigma_j} \right)^2
  \label{eq:Gamma}
\end{equation}
where $\eta$ represents the effects of nonuniform ionic diameters
\begin{equation}
  \eta = \frac{1}{\Omega} \frac{\pi}{2\Delta} \sum_j \frac{\rho^{bath}_j \sigma_j z_j}{1 + \Gamma \sigma_j}
\end{equation}
and $\Omega$ is determined by
\begin{equation}
  \Omega = 1 + \frac{\pi}{2\Delta} \sum_j \frac{\rho^{bath}_j \sigma^3_j}{1 + \Gamma \sigma_j}
\end{equation}
This implicit relationship is a quartic equation in $\Gamma$, which we solve using Newton's method. We may then
calculate the bath potential
\begin{equation}
  \mu^{ES,bath}_i = -\frac{e^2}{4\pi\epsilon\epsilon_0} \left[ \frac{\Gamma z^2_i}{1 + \Gamma\sigma_i} +
    \eta\sigma_i \left( \frac{2z_i - \eta\sigma^2_i}{1 + \Gamma\sigma_i} + \frac{\eta\sigma^2_i}{3} \right) \right].
\end{equation}

\section{Evaluation of the Fourier Transform of the Weighting Functions \label{app:Fourier-of-weights}}
%\comment{Is the following paragraph really necessary?  It is rather particular to FFTW, which is not being said here.}
We must be careful to evaluate our analytic transforms at the same $\veck$ values, in the same order, as those computed
using the particular implementation of FFT we use. Given a $D$ dimensional grid, the vector $\{k_d\}$ which corresponds
to the vertex $\{j_d\}$ of our Cartesian grid is given by
\begin{equation}
  k_d = \begin{cases}
    \frac{ 2\pi j_d}{N_d h_d}         & j_d \leq \frac{N_d}{2} \\
    \frac{-2\pi (N_d - j_d)}{N_d h_d} & j_d >    \frac{N_d}{2}
  \end{cases}
  \label{eq:Kvalues}
\end{equation}
where $N_d$ is the number of grid points in dimension $d \in \{x,y,z\}$, and $h_d$ is the grid spacing
$\frac{L_d}{N_d-1}$.

We begin with the calculation of $\hat{\omega}^2_i$,
\begin{align}
  \hat{\omega}^2_i &= \int^{2\pi}_0 d\phi \int^\pi_{0} d\theta\sin\theta \int^\infty_0 dr\ r^2
    \delta\left(|r| - R_i\right) e^{-\imath\vec{k}\cdot\vec{x}} \\
                   &= \int^{2\pi}_0 d\phi \int^\pi_{0} d\theta\sin\theta R^2_i e^{-\imath R_i\vec{k}\cdot\hat{x}}.
\end{align}
We now choose a rotated coordinate system (the prime system) in which $\vec{k}$ points purely in the $z'$ direction, in
order to take advantage of the rotational symmetry of the problem. In the new coordinate system,
\begin{eqnarray}
  \hat{\omega}^2_i &=& \int^{2\pi}_0 d\phi' \int^\pi_0 d\theta' \sin\theta' R^2_i e^{-\imath R_i k'_z\cos\theta'} \\
                   &=& 2\pi R^2_i \int^\pi_0 d\theta' \sin\theta' \left(\cos\left(R_i k'_z\cos\theta'\right) -
                       \imath \sin\left(R_i k'_z\cos\theta'\right)\right) \\
                   &=& \frac{4\pi R_i \sin\left(R_i k'_z\right)}{k'_z}
\end{eqnarray}
which, in the original coordinate system, is
\begin{equation}
  \hat{\omega}^2_i = \frac{4\pi R_i\sin\left(R_i |\vec{k}|\right)}{|\vec{k}|}.
  \label{eq:OmegaHat2}
\end{equation}
From Eq.~(\ref{eq:omegaDef}), we also have
\begin{equation}
  \hat{\omega}^0_i = \frac{\sin\left(R_i |\vec{k}|\right)}{R_i |\vec{k}|} \qquad
    \hat{\omega}^1_i = \frac{\sin\left(R_i |\vec{k}|\right)}{|\vec{k}|}. \label{eq:omegaHat0and1}
\end{equation}
Recognizing that the theta function can be obtained as the integral of a delta function, we have
\begin{align}
  \hat{\omega}^3_i &= \int^{R_i}_0 dr\, \hat{\omega}^2_i \mid_{R_i = r} \\
                   &= \frac{4\pi}{|\vec{k}|} \int^{R_i}_0 dr\ r \sin\left(r |\vec{k}|\right) \\
                   &= \frac{4\pi}{|\vec{k}|^3} \left(\sin\left(R_i |\vec{k}|\right) - R_i |\vec{k}|\cos\left(R_i |\vec{k}|\right)\right) \label{eq:omegaHat3}
\end{align}
Following a similar procedure as in the $\hat{\omega}^2_i$ calculation, but keeping track of the vector nature of
$\omega^{V1}$ and $\omega^{V2}$,
\begin{eqnarray}
  \hat{\omega}^{V2}_i &=& \int^{2\pi}_0 d\phi' \int^\pi_0 d\theta' \sin\theta' R_i^{2} e^{-\imath R_i k'_z\cos\theta \hat{x}} \\
                      &=& -2\pi\imath R_i \int^\pi_0 d\theta' \sin\theta' \cos\theta' \sin\left(R_i k'_z\cos\theta'\right) \hat{k'_z} \\
                      &=& \frac{-4\pi\imath}{|\vec{k}|^{2}}\left(\sin\left(R_i |\vec{k}|\right)
                          - R_i |\vec{k}|\cos\left(R_i |\vec{k}|\right)\right) \hat{k}.
  \label{eq:omegaHatV2}
\end{eqnarray}

The preceding expressions for $\hat{\omega}^\alpha_i$ may be evaluated at $|\vec{k}| = 0$, but care must be taken when calculating the limit.
\begin{align*}
  \underset{|\vec{k}|\rightarrow0}{\lim} \hat{\omega}^0_i &= \underset{|\vec{k}|\rightarrow0}{\lim} \frac{\sin\left(R_i|\vec{k}|\right)}{R_i|\vec{k}|}
    = 1 \\
  \underset{|\vec{k}|\rightarrow0}{\lim} \hat{\omega}^1_i &= R_i \\
  \underset{|\vec{k}|\rightarrow0}{\lim} \hat{\omega}^2_i &= 4\pi R_i^{2} \\
  \underset{|\vec{k}|\rightarrow0}{\lim} \hat{\omega}^3_i &= \underset{|\vec{k}|\rightarrow0}{\lim}
    \frac{4\pi}{|\vec{k}|^{3}} \left( \left(R_i |\vec{k}| - \frac{\left(R_i |\vec{k}|\right)^{3}}{6}\right) -
    R_i| \vec{k}| \left(1 - \frac{\left(R_i| \vec{k}|\right)^{2}}{2}\right)\right) \\
  &= \frac{4}{3}\pi R_i^{3} \\
  \underset{|\vec{k}|\rightarrow0}{\lim} \hat{\omega}^{V1}_i &= 0 \\
  \underset{|\vec{k}|\rightarrow0}{\lim} \hat{\omega}^{V2}_i &= 0
\end{align*}
It should be noted these are the limits one would expect, since in the $|\vec{k}|=0$ case we are simply integrating
either a spherical delta or step function over all space, thereby recovering surface area and volume expressions for a
sphere.

\section{Directional Average Bound \label{app:average-bound}}

We can bound the directional average of the density over a sphere in terms of the unweighted average, and thus we can
bound the ratio
\begin{equation}
  \frac{\left| n^{V2}(x) \right|^2}{\left| n^2(x) \right|^2}
\end{equation}
in the calculation of $\Phi(n)$ from Eq.~\ref{eq:Phi}. We let $\nu(\theta,\phi)$ be the unit vector at the surface of
the sphere in the $(\theta,\phi)$ direction. Using Fubini's Theorem and the Cauchy-Schwarz inequality, we have
\begin{eqnarray}
  n^2_{V2}(x) &=&   \sum_{ij} \int_{S^2} \nu(\theta,\phi) \rho_i(x + r) d\Omega \cdot \int_{S^2} \nu(\theta',\phi') \rho_j(x + r') d\Omega' \\
              &=&   \sum_{ij} \int_{S^2\times S^2} \nu(\theta,\phi) \cdot \nu(\theta',\phi') \rho_i(x + r) \rho_j(x + r') d\Omega d\Omega' \\
              &\le& \sum_{ij} \int_{S^2\times S^2} |\nu(\theta,\phi) \cdot \nu(\theta',\phi')| |\rho_i(x + r) \rho_j(x + r')| d\Omega d\Omega' \\
              &\le& \sum_{ij} \int_{S^2\times S^2} |\nu(\theta,\phi)| |\nu(\theta',\phi')| \rho_i(x + r) \rho_j(x + r') d\Omega d\Omega' \\
              &\le& \sum_{ij} \int_{S^2\times S^2} \rho_i(x + r) \rho_j(x + r') d\Omega d\Omega' \\
              &=&   \sum_{ij} \int_{S^2} \rho_i(x + r) d\Omega \int_{S^2} \rho_j(x + r') d\Omega' \\
              &=&   n^2_2(x)
\end{eqnarray}
so that
\begin{equation}
  \frac{\left| n^{V2}(x) \right|^2}{\left| n^2(x) \right|^2} \le 1.
  \label{eq:ratio}
\end{equation}

%\comment{DMITRY: Insert analysis of roundoff error for the ratio here!}

%\bibliographystyle{plain}
\bibliography{petscapp,Writeup}

\end{document}